\documentclass[prc,reprint]{revtex4-1}

\usepackage{graphicx}
\usepackage{hyperref}
\usepackage{wasysym}   

\begin{document}

\title{Observation of hard radiations in a laboratory atmospheric high-voltage discharge} 

\author{A.V.~Agafonov}
\affiliation{P.N.~Lebedev Physical Institute of the Russian Academy of Sciences (FIAN), Moscow, 119991, Leninsky pr., 53}
\author{V.A.~Bogachenkov}
\affiliation{P.N.~Lebedev Physical Institute of the Russian Academy of Sciences (FIAN), Moscow, 119991, Leninsky pr., 53}
\author{A.P.~Chubenko}
\affiliation{P.N.~Lebedev Physical Institute of the Russian Academy of Sciences (FIAN), Moscow, 119991, Leninsky pr., 53}
\author{A.V.~Oginov}
\email{oginov@lebedev.ru} \affiliation{P.N.~Lebedev Physical
Institute of the Russian Academy of Sciences (FIAN), Moscow,
119991, Leninsky pr., 53}
\author{A.A.~Rodionov}
\affiliation{P.N.~Lebedev Physical Institute of the Russian Academy of Sciences (FIAN), Moscow, 119991, Leninsky pr., 53}
\author{A.S.~Rusetskiy}
\affiliation{P.N.~Lebedev Physical Institute of the Russian Academy of Sciences (FIAN), Moscow, 119991, Leninsky pr., 53}
\author{V.A.~Ryabov}
\affiliation{P.N.~Lebedev Physical Institute of the Russian Academy of Sciences (FIAN), Moscow, 119991, Leninsky pr., 53}
\author{A.L.~Shepetov}
\affiliation{P.N.~Lebedev Physical Institute of the Russian Academy of Sciences (FIAN), Moscow, 119991, Leninsky pr., 53}
\author{K.V.~Shpakov}
\affiliation{P.N.~Lebedev Physical Institute of the Russian Academy of Sciences (FIAN), Moscow, 119991, Leninsky pr., 53}

\date{\today}

\begin{abstract}

The new results concerning neutron emission detection from a
laboratory high-voltage discharge in the air are presented. Data
were obtained with a combination of plastic scintillation
detectors and $^3$He filled counters of thermal neutrons. Strong
dependence of the hard x-ray and neutron radiation appearance on
the field strength near electrodes, which is determined by their
form, was found. We have revealed a more sophisticated temporal
structure of the neutron bursts observed during of electric
discharge. This may indicate different mechanisms for generating
penetrating radiation at the time formation and development of the
atmospheric discharge.

\end{abstract}


\maketitle

\section{Introduction}

The high-energy emissions of electrons, gamma rays, and neutrons
produced in association with lightning and thunderclouds have been
observed in a number of space and ground-based experiments.
Elucidation of the mechanisms of high-energy radiation arising at
the time of atmospheric discharge is an increasingly expanding
area, both of experimental and theoretical research.  A
comprehensive review of the physical phenomena occurring in the
thunderstorm atmosphere is presented in~\cite{dwy} where this
field of study comes under designation of ``high-energy
atmospheric physics''.

Of particular interest are occasional observations of temporary
enhancement of the neutron flux inside thunderstorm atmosphere.
The first evidences of a surplus neutron radiation at thunderstorm
time have been reported in~\cite{sha}. In recent years, some
papers discuss the possibility of neutron generation in connection
with powerful lightning discharges on the basis of experimental
observations made at the mountain heights at
Tien~Shan~\cite{gur,gur2}, Aragats~\cite{chi,chi2},~Tibet
\cite{tsu}, so as at the sea level in MSU~\cite{kuzh} and
Yakutsk~\cite{koz}. Up to date, the different physical mechanisms
were being discussed for explanation of thunderstorm related
neutron production, but the nature this phenomenon still remains
unknown.

Recently we reported our first measurements of neutron emission
originated by atmospheric lightning like discharge in a laboratory
installation~\cite{aga}. Experiments were carried out inside an
electric field with the average strength of the order of
$\sim$1~MV$\cdot$m$^{-1}$ and with neutron detectors of two
independent kinds: the CR-39 type tracers and the plastic
scintillators. The neutrons were registered within the range from
thermal energies up to the energies above 10~MeV. It was found
that the neutron generation takes place at the initial phase of
electric discharge and is correlated with generation of x-ray
radiation. The estimations of the average density of neutron flux
lay in the range of (0.2--1)$\times$10$^6$~cm$^{-2}$ per a
discharge shot. We did not found any convincing explanation for
observed high values of neutron flux according to the registration
in the CR-39 track detectors. A detailed theoretical analysis of
various nuclear reactions which could contribute to this flux is
presented in \cite{bab}, and it ends with conclusion that the
``known fundamental interactions cannot allow prescribing the
observed events to neutrons''.

Insufficient knowledge on physical processes which could take
place at the initial stage of the atmospheric discharge and the
complexity of any self-consistent numerical simulation of this
phenomenon determines the need for a comprehensive experimental
study of the formation of discharge and diagnosis of different
accompanying radiations. The use of a laboratory installation with
its parameters sufficient for a precise temporal and spatial
separation of running processes allows to establish general laws
and to check expected hypothesis on discharge formation. In
present article we report the new results concerning detection of
neutron emission from a controlled electric discharge in the air
which was made with a combination of plastic scintillation
detectors and $^3$He filled proportional ionization counters of
thermal neutrons. This study has revealed a more sophisticated
temporal structure of the neutron bursts observed at the time of
electric discharge. It was found that the flash of hard x-rays
generated by discharge is not always accompanied by any appearance
of neutron signal as well as both the single and multiple neutron
pulses were recorded which do not overlap with any x-ray pulses at
all. In some rather rare events the neutron pulses appear at the
final stage of discharge, when the voltage has just fell down to
zero, or even after polarity change of accelerating voltage, and
do not correlate with any x-ray radiation.

\section{Experimental setup, detector calibration, and the time structure of radiation pulses}

\begin{figure}
\resizebox{0.5\textwidth}{!}{
\includegraphics{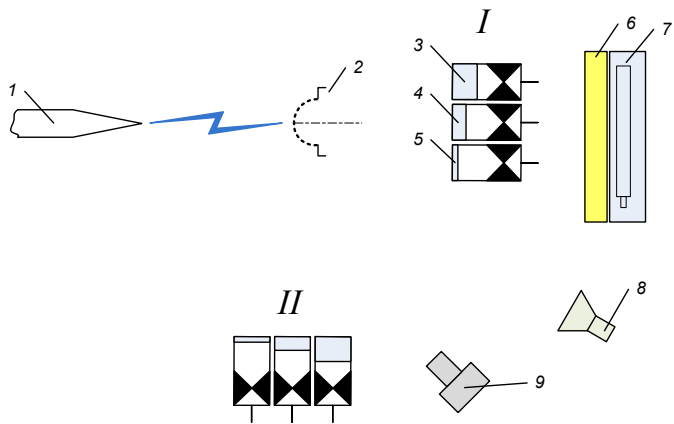}}
\caption{ \label{fig:1} Schematic of apparatus layout in laboratory experiment: {\it 1} --- cathode, {\it 2} --- anode, {\it 3} --- SD1 (10~cm Pb), {\it 4} --- SD2 (6~mm Pb), {\it 5} --- SD3 (50~$\mu$m Al), {\it 6} --- paraffin moderator, {\it 7} --- array of $^3$He-counters, {\it 8} --- RF horn antenna, {\it 9} --- lateral camera. \textit{I} and \textit{II} --- different positions of the triple scintillation detectors SD1, SD2, SD3.}
\end{figure}

Experiments were held with a high current electron accelerator ERG
destined for investigation of the high-voltage discharge in the
air~\cite{n1}. The main characteristics of generated discharge and
procedure of electrophysical diagnostics were the same as in the
measurements described in \cite{aga}. High voltage pulses with a
$\sim$1~MV amplitude were applied to a discharge air gap of the
450--750~mm width (the latter was changed in various runs). The
current pulse amplitude was about 10--12~kA, and the total
duration of pulses was about $\sim$350--1000~ns, in dependence on
the gap width. In successive experimental series were used the
electrodes of different configurations: the cathode and anode of
hemispherical shape, 80~mm and 90~mm in diameter, and the
semispherical $\diameter$90~mm mesh anode in combination with
cathode needle. The scheme of the current experimental set-up with
a set of plastic scintillation detectors, the $^3$He filled
thermal neutron counters, optic sensors, and radio antenna is
shown in the Fig.~\ref{fig:1}.

In both experimental series one and the same assembly of three
plastic scintillation detectors (SD1--SD3) was used which is shown
in positions {\it I} and {\it II} in the Fig.~\ref{fig:1}. All
detectors are build on the basis of $15\times 15\times 5$~cm$^3$
scintillator blocks optically coupled with the FEU30 type
photomultiplier tube (PMT); the detector SD1 is shielded with a
10~cm thick wall of the stacked lead bricks from all its sides
(even at 100-fold attenuation of gamma radiation threshold energy
it is approximately E$_{\gamma} \approx 2$~MeV), SD2 is wrapped by
the two layers of sheet lead with sum thickness of 6~mm (which
corresponds to a gamma rays cutoff energy E$_{\gamma}\approx
200$~keV), and SD3 is covered by an aluminum foil with the
thickness of 50~$\mu$m (cutoff energy E$_{\gamma}\approx 10$~keV).

Location {\it I} of scintillation assembly was used in the most
part of experiments. In this position, the SD1 detector was placed
at the distance of 100~mm from the axis of the system, and 470~mm
behind the anode; SD2 just on the axis and at the distance of
470~mm from the anode; and SD3 at the distance of 880~mm from
anode and 110~mm aside the axis, and below the level of SD1 by the
values of its height. The position {\it II} was used to measure
the weakening of radiation intensity and to estimate its energy in
supposition if all generated emission is considered as gamma rays.
In second position all SD were located in one vertical plane
900~mm aside of discharge axis, and the detector SD3 was placed
between SD1 and SD2.

All three scintillation detectors were calibrated with a standard
x-ray source RINA~\cite{n2} and have approximately one and the
same sensitivity relative to electromagnetic radiation. X-ray tube
of IMA6 D type powered by 0.15~J pulses with peak voltage of
100~kV. Angular divergence of x-rays is 30$^\circ$, the pulse
duration is about 10~ns with repetition rate of 8~Hz. The diameter
of the effective focal spot is 2.5~mm.

For continuous (not pulsed) mode estimation their registration
efficiency relative to neutron flux the scintillation detectors
were irradiated by a $^{252}$Cf neutron source with overall yield
of $3\times10^4$~neutrons per second. In the case where the source
is placed in the center point of the scintillator plane the
detection efficiency of the detector was 0.17. When the
scintillation detector was placed inside a 10-cm lead box and the
source was at the anode (at a distance of about 0.5~m)
scintillator cannot record signal (background level) because of
small average intensity of neutron flux from the source.

As an alternative and independent method of neutron registration
in the present experiment was applied the multichannel detector on
ionization neutron counters. This detector is based on the
$^3$He-filled neutron counters of SNM-18 type which were operating
in proportional mode. Twelve $\diameter3$cm$\times30$~cm counters
together with all necessary electronics were put into a 2~mm thick
duraluminium casing covered by the discharge (front) side with a
7~cm thick paraffin layer for moderation of anticipated neutrons.
The total area of the neutron detection block is 1000~cm$^2$; the
counters are filled with the pure $^3$He gas under the pressure of
2~atm. When operating, the detector was placed in position at
70~cm from the anode and 130~cm from the cathode as shown in
Fig.~\ref{fig:1}, with its front moderator cover being turned
against the generator. The pulses from all neutron counters were
recorded separately, so the average intensity of neutron flux
could be calculated as a sum of these counts.

When using the ionization counters based detectors in the vicinity
of a powerful electrical installation, it is a general problem to
suppress the strong electromagnetic interference on their counts
from the nearby electric discharges. Therefore, the neutron
counters together with their signal acquisition electronics were
especially locked inside an electrically shielded volume. Also, in
a part of experiments the whole counter assembly was placed inside
an additional tight box of welded 3~mm thick iron, and in this
position another 43~mm thick layer of light material (perplex) ---
a neutron reflector --- was put against the back (opposite to
discharge generator) side of the counter case.

Absolute calibration of neutron detector was made experimentally
with the use of $^{252}$Cf neutron source. With the source placed
in vicinity of cathode the net neutron detection efficiency of
described $^3$He counters assembly occurred to be 0.076\%, and
with the source in anode region --- 0.12\%.

\begin{figure}
\resizebox{0.3\textwidth}{!}{\includegraphics{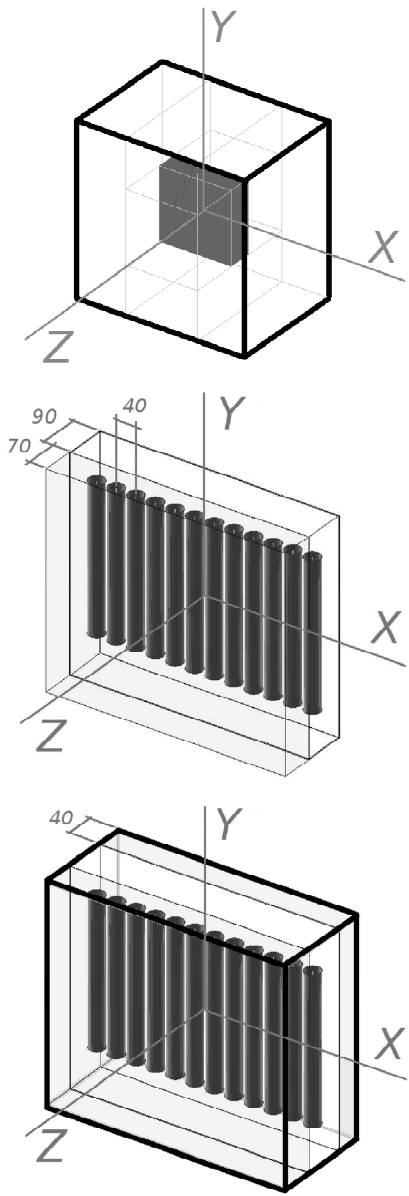}}

\caption{Three neutron detector models used in Geant4 simulation:
the scintillator block inside a cube of 10~cm thick lead walls
(model~{\it I}); the set of SNM-18 neutron counters with a front
side moderator (model~{\it II}); the same counter assembly with
addition of a back side reflector placed into a solid iron box
(model~{\it III}). Front side (where the primary particles were
coming from) always corresponds to positive direction of {\it Z}
axis, dimensions are shown in mm.} \label{geantmodels}
\end{figure}

Besides these experimental measurements, efficiency of neutron
registration was defined in complete simulation of the neutron
propagation process both inside the neutron detectors and in
surrounding materials with the use of Geant4
toolkit~\cite{geant4}. For the purpose, three detector models were
built which took into account specific features of detector
set-ups used in reality (see the Fig.~\ref{geantmodels}).

The model~{\it I} corresponds to the typical configuration of SD1
scintillation detector: a $15\times 15\times 5$~cm$^3$ plastic
scintillator block surrounded by 10 cm thick lead walls. In the
model~{\it II} a set of 12 $^3$He filled cylindrical ``counters''
having the size of a real SNM-18 type detector (3~cm in diameter
and 30~cm long) were put into a box of Al material with a 7~cm
thick paraffin moderator layer placed at its front side. In
model~{\it III} this counter assembly was supplemented with a
backward neutron reflector of 4~cm thick perspex, and the whole
set-up was placed inside a solid metallic box of a 3~mm thick iron
sheets.

The particle physics module of simulation program took into
consideration the following processes of neutron interaction: the
Geant4 models of elastic coincidences in the range from thermal
energies (of the order of $10^{-2}$~eV) up to 4~eV, so as the
elastic coincidence models of the intermediate (4~eV--20~MeV) and
high-energy neutrons (above 20~MeV); the models of inelastic
interaction in the ranges of thermal, intermediate and high
energies; the models of radiative neutron capture. For the
protons, besides analogous processes of their elastic and
inelastic interactions the models of multiple scattering and
ionization losses were considered; and for the positive and
negative pions the decay process (the physics of negative pions
included also the process of their absorption at rest). The fact
of neutron registration was signalled either by appearance of any
charged particle (mostly, a recoil proton) with the energy above
0.5~MeV inside the scintillator volume of the model~{\it I}, or by
a $^3$H nucleus born inside the volume of any neutron counter for
the case of models~{\it II} and~{\it III} (correspondingly to
nuclear reaction $n+^3$He~$\rightarrow p+^3$H which is used in the
real SNM-18 type detectors). In turn, the overall efficiency of
neutron registration was defined as a relation of the number of
``registered'' neutrons to the total number of primary particles
put into simulation.

\begin{figure}
\resizebox{0.4\textwidth}{!}{\includegraphics{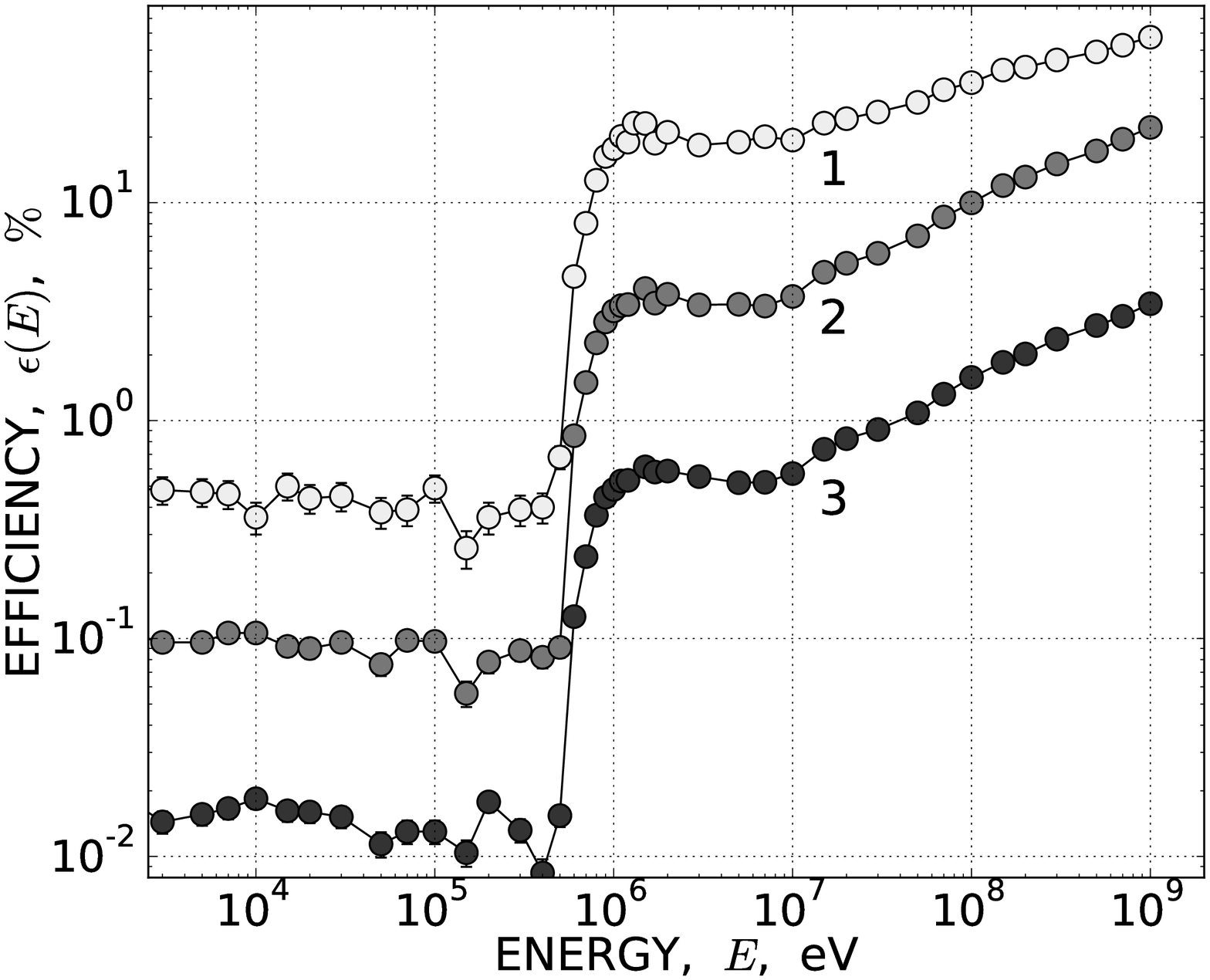}}
\resizebox{0.4\textwidth}{!}{\includegraphics{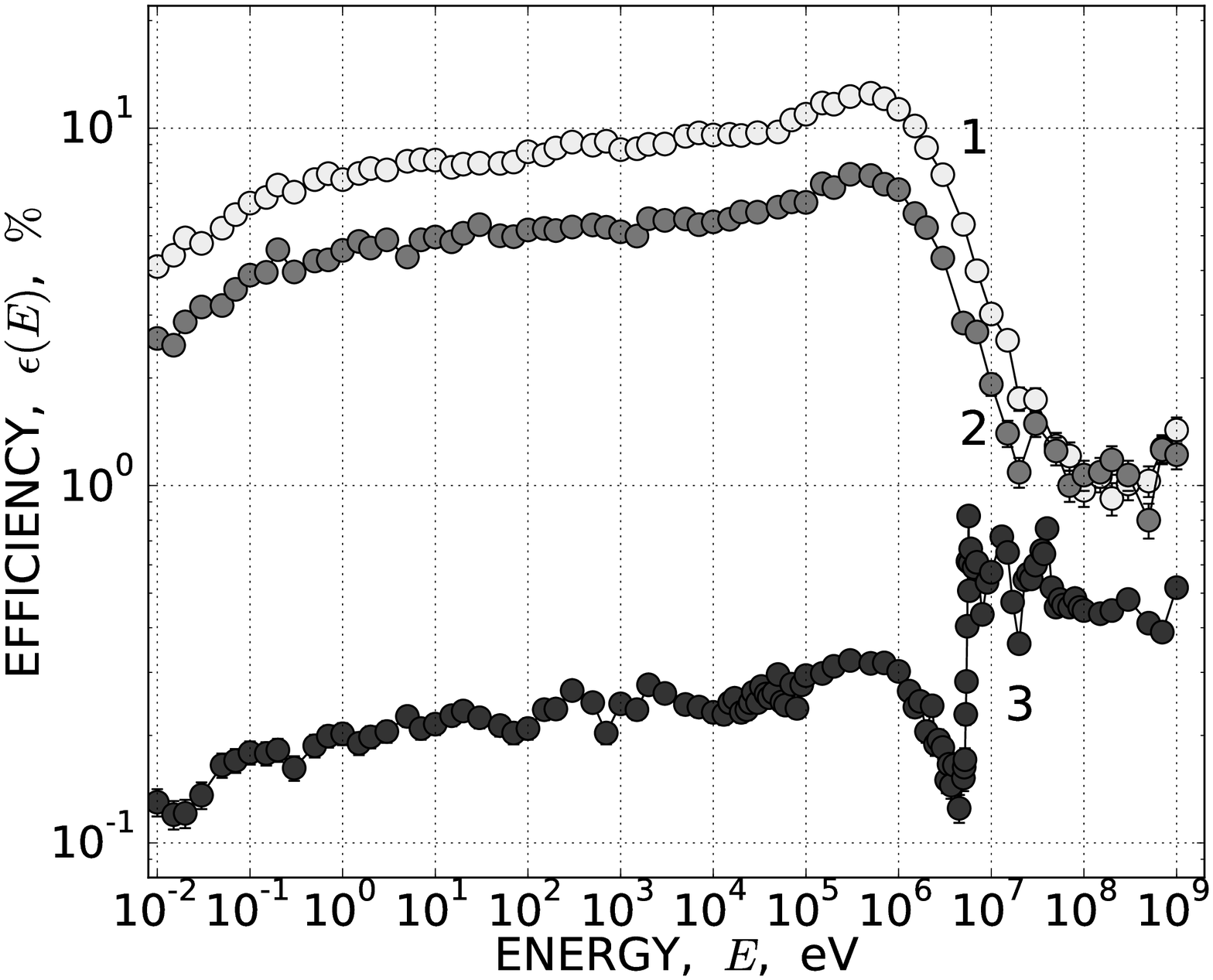}}
\resizebox{0.4\textwidth}{!}{\includegraphics{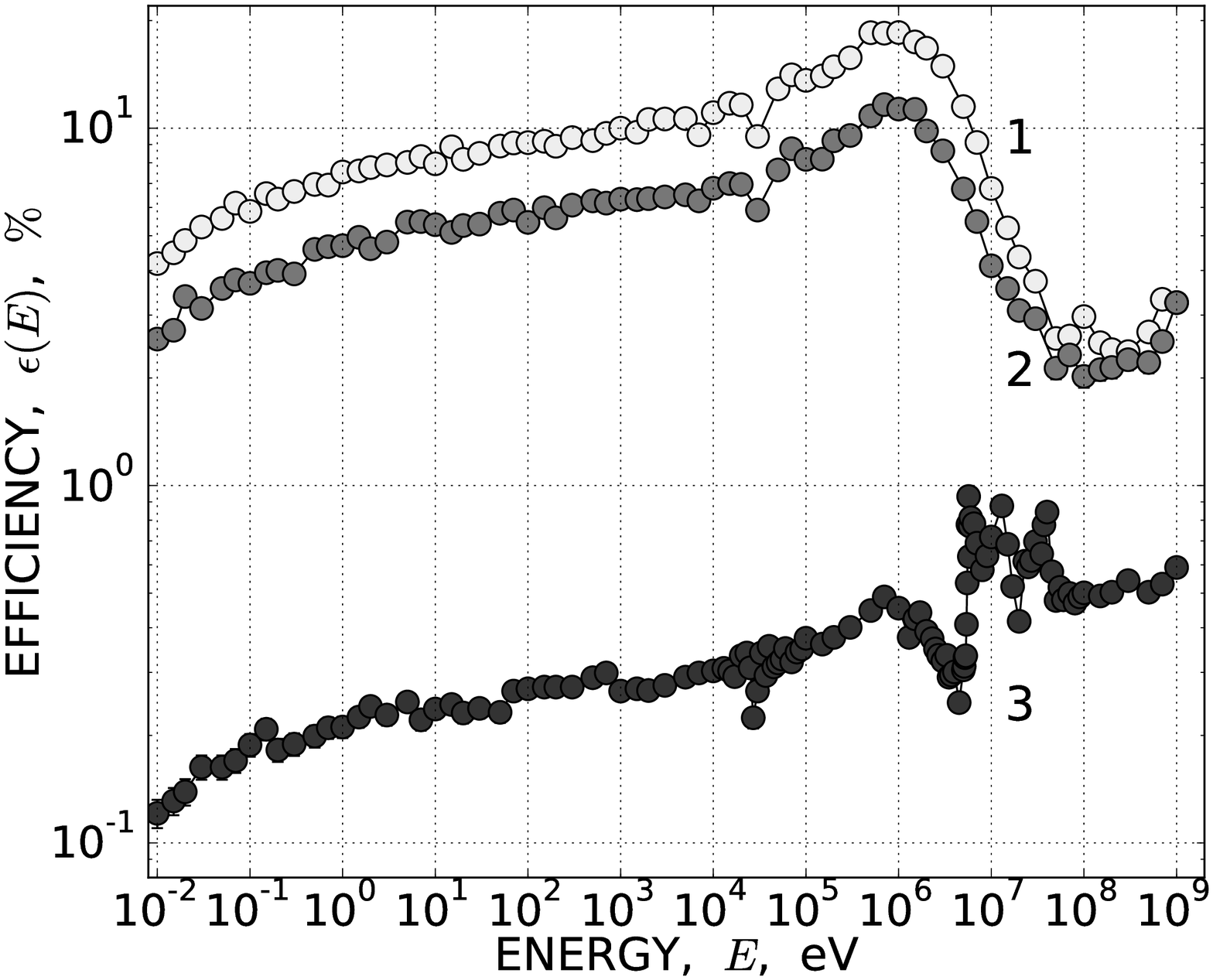}}
\caption{Efficiency of neutron registration with different detector types used in present experiment (result of Geant4 simulations). From top to bottom frames: the scintillation neutron detector inside a lead shielding assembly (detector model~{\it I}); the set of ionization neutron counters with 7~cm thick neutron moderator before its front side (model~{\it II}); the same set of counters together with its front side moderator and back side reflector put in an iron box (model~{\it III}). Curves~{\it 1} correspond to simulation for a normal beam of monoenergetic primary neutrons hitting the center of detector front side (ideal case); {\it 2} --- for the neutrons distributed randomly over detector front with their momenta perpendicular to the surface;  {\it 3} --- for a point-like isotropic source of primary neutrons displaced 70cm apart from the front surface (realistic case).}
\label{geantresults}
\end{figure}

Three distribution variants of primary neutron particles were
accepted in simulation series for every detector model~{\it I},
{\it II}, and~{\it III}. In first turn, a parallel beam of
monoenergetic primary neutrons was falling on center point of
detector front perpendicularly to its surface (an ideal case of
primary geometry and upper limit of possible registration
efficiency). In second (intermediate) variant the position of
primary neutrons was randomly selected on the front surface while
their momenta remain always perpendicular to it. In the third
(realistic) simulation the primary neutrons were emitted with
isotropic direction distribution from a spherical point source
displaced 70~cm apart from the front surface, similarly to
geometry of real experiment were origination of neutrons is
supposed to have place somewhere around electrodes of high voltage
generator. A number of succeeding simulation runs with constantly
increasing energies of primaries was fulfilled for every
combination ``detector model/primary geometry'', and a set of
registration efficiency distributions was calculated in dependence
on neutron energy. These results are presented in the
Fig.~\ref{geantresults}.

As is known, the energy spectrum of neutron emission from
$^{252}$Cf isotope fission does spread up to 13~MeV with its mean
value being 2.3~MeV and the most probable value 1.2~MeV. According
to the upper frame of Fig.~\ref{geantresults} the maximum average
efficiency of neutron registration by scintillation detector in
the energy range of some MeV must be confined by the curves~{\it
1} and~{\it 2}, i.e. somewhere between 5--20\%. For the
``realistic'' configuration of a distant isotropic neutron source
the curve~{\it 3} predicts an efficiency about 0.4--0.5\%.

The simulation results made for a ionization neutron counter based
detector in the middle and bottom frames of
Fig.~\ref{geantresults} have rather irregular behaviour in MeV
energy range for the case of a distant point source (the
curve~{\it 3}). This can be explained through manifestation of an
additional neutron multiplication mechanism due to recoil process
in surrounding materials around the neutron counters; this effect
starts to be feasible just around the MeV energy threshold and
occurs being noticeable in the curves~{\it 3} because of their
generally low efficiency level.

In the case of the neutron detector model~{\it II} the simulation
predicts the mean registration efficiency about 0.15-0.20\% for a
distant primary source (curve~{\it 3}) which is close to 0.12\%
efficiency measured with location of californium source in anode
region. The presence of additional backward neutron reflector in
the case of model~{\it III} reveals itself through the increase of
resulting registration efficiency up to the values of 0.30-0.45\%.

\begin{figure}
\resizebox{0.49\textwidth}{!}{
\includegraphics{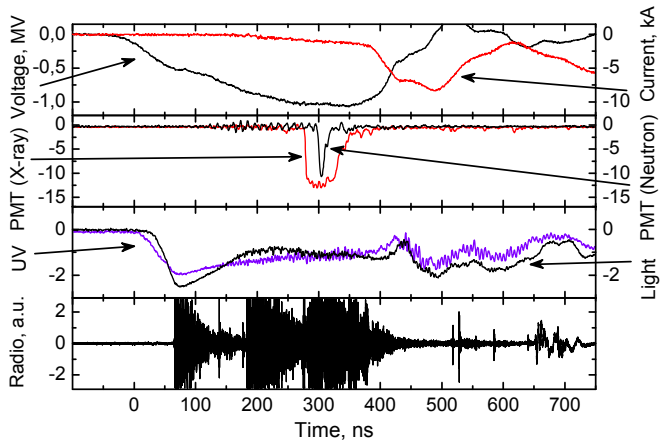}}
\caption{Real-time observation: the oscilloscope traces of the
current, voltage, x-ray, and pulses of neutron signal on the PMT
output of the scintillation neutron detector. Also shown the
waveforms of accompanying optic and ultraviolet radiation, and of
the radio-emission.  The x-ray, neutron, optic, ultraviolet, and
radio-emission signal amplitudes are presented in relative units.}
\label{fig:3}
\end{figure}

Discharge development in the real-time experiment was controlled
by its photographing at different angles with the use of
multichannel digital photo-recorder and application of the neutral
and colored optical glass filters, and trough the fast
oscilloscope recording of characteristic electric parameters and
radiation intensities. The Fig.~\ref{fig:3} presents the typical
time behavior of the electric current and voltage on the discharge
gap, so as the intensity of 1--3~GHz radio-emission, optic light,
ultraviolet and x-ray radiation, and neutron pulse. It is seen
that the moment of x-ray pulse usually is delayed up to
200--300~ns relative to the voltage jump, and coincides with the
time of increasing pre-pulse current; most often, the neutron
signals do occur within the time of the x-ray radiation pulse.

\section{The real-time neutron and x-ray measurements}

The lightning discharges are recorded simultaneously with neutrons
and gamma radiation using scintillation detectors.  Detection of
neutrons and/or hard gamma rays in a laboratory experiment
requires a clear analysis.

Firstly, we have identified a strong dependence of the appearance
of the hard (neutron or gamma) radiation from the electric field
strength near the electrodes, which is determined by their form.
Statistics of events with the advent of x-rays and neutrons are
presented in Table~\ref{tab1}.

\begin{table*}
\caption{\label{tab1}Statistics of events with the advent of
x-rays and neutrons.}
\begin{ruledtabular}
\begin{tabular}{cccc}
 Geometry of the electrodes& Shots with x-ray& Shots with
 x-ray& Shots with\\
 and the total number of shots&($>10$~keV)(SD2)&($>100$~keV)(SD3)&neutron pulses (SD1)\\ \hline

Hemishpere (cathode)-&&& \\
hemishpere (anode)&&& \\
 341&39\%&33\%&1.4\% \\
including ``cut-off''&&& \\
96&36\%&35\%&2\% \\
\\ \hline
Needle (cathode)-&&& \\
hemishpere (anode)&&& \\
 950&100\%&87\%&15\% \\
including ``cut-off''&&& \\
20&100\%&92\%&15\% \\
\\
\end{tabular}
\end{ruledtabular}
\end{table*}

It is necessary underline very large amplitudes spread of signals
registered by SD1. Although 15\% is indicated in second line of
Table~\ref{tab1}, only 5\% of pulses have its amplitude exceeding
a third of maximum one.

Secondly, we investigated in detail the temporal structure of the
appearance of neutron pulses. In a previous paper~\cite{aga} we
noted that the momentum of the neutron radiation, the detected
scintillation detector placed behind a lead shield, is correlated
with the occurrence of x-ray pulse and is located within it. That
is, the neutron pulse was observed in the initial ``dark'' phase
of the discharge until closure of cathode and anode streamers.
However, the structure of the neutron pulse was more complicated.

\begin{figure*}
\resizebox{0.95\textwidth}{!}{
\includegraphics{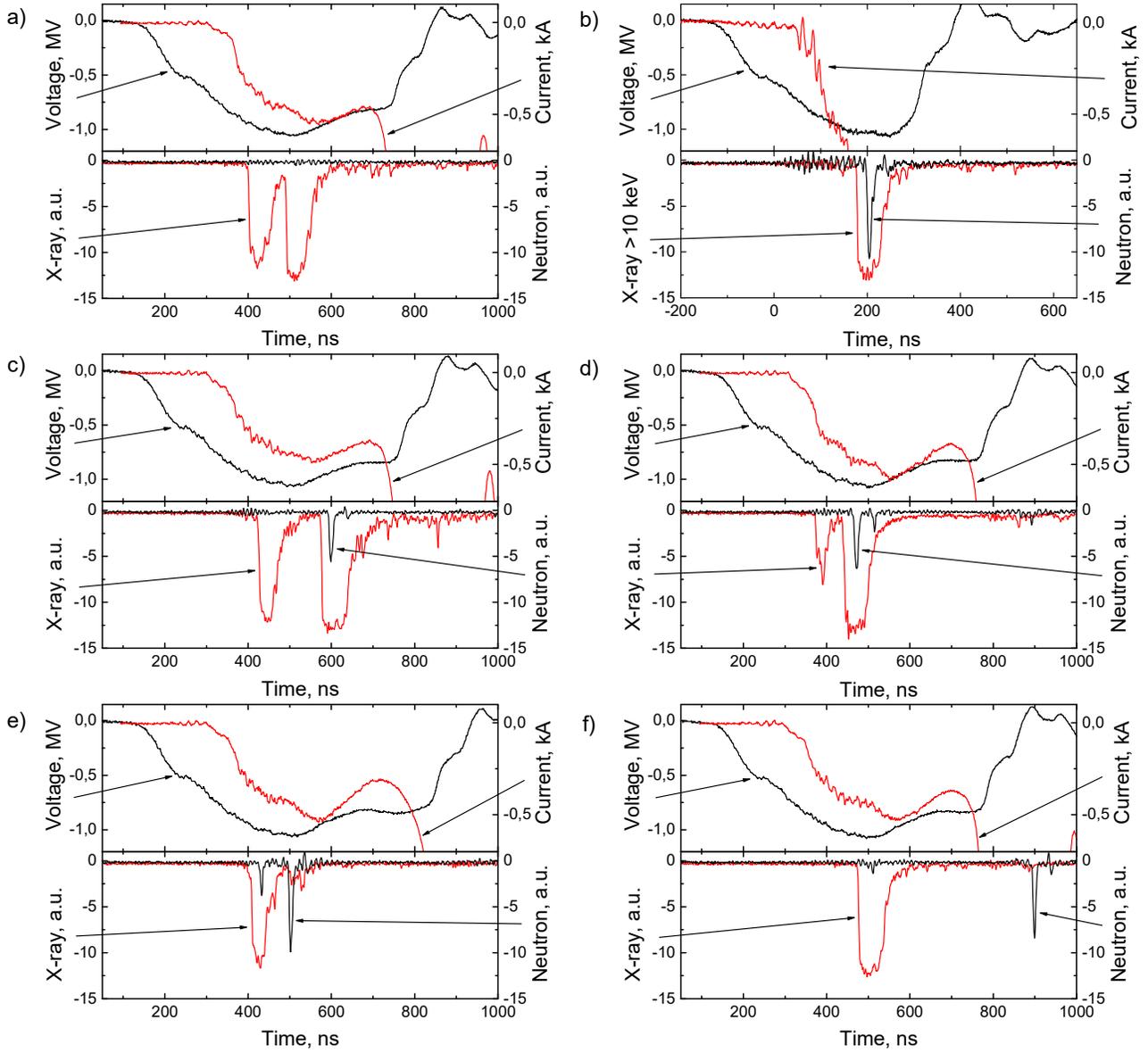}}
\caption{X-rays and neutron radiation pulses relative to voltage
and pre-pulse current traces.} \label{fig:4}
\end{figure*}

New measurements have shown that the structure and characteristics
relative arrangement pulses of x-rays and neutron radiation has
been quite diverse (Fig.~\ref{fig:4}). The emergence of x-ray
radiation is not always accompanied by the appearance of neutrons
(Fig.~\ref{fig:4}a). The most common occurrence of neutron pulses
clearly correlated with the x-rays pulse and are located inside
(Fig.~\ref{fig:4}b). Recorded both single and binary neutron
pulses inside x-ray(Fig.~\ref{fig:4}c,d), and near the x-ray
pulses (Fig.~\ref{fig:4}e). In some rare cases, the neutron pulse
appears at the final stage of the discharge (Fig.~\ref{fig:4}f) in
the absence of x-ray radiation. In most cases, the neutron
impulses occur near the peak voltage to the main phase of the
discharge (at the peak of pre-pulse current).

\begin{figure*}
\resizebox{0.95\textwidth}{!}{
\includegraphics{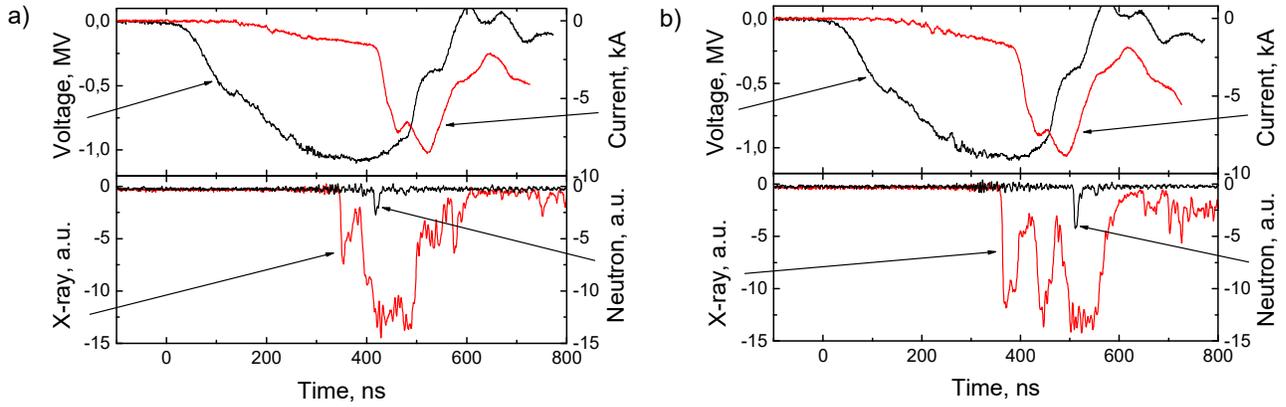}}
\caption{X-rays and neutron radiation pulses relative to voltage
and total current traces.} \label{fig:5}
\end{figure*}

However, there are cases where the neutron pulse is generated at
the beginning of the main phase (Fig.~\ref{fig:5}a) or at the peak
of the discharge current and correlated with the x-ray pulse
(Fig.~\ref{fig:5}b). In Fig.~\ref{fig:5} unlike Fig.~\ref{fig:4}
instead of initial part (pre-pulse) total discharge current is
shown.

Various provisions of the neutron radiation pulse relative to the
pulse voltage and discharge current leads to the assumption of
possible different mechanisms of neutron generation in the initial
stage of discharge (until the end of streamer-leader stage) and
its main stage.

The amplitude of the neutron signals in a variety of shots can
vary by an order of magnitude. Taking into account the efficiency
of the neutron detection by the scintillation detector and under
the conditions that the flow is isotropic spectrum and the energy
does not differ greatly from the spectrum of the source
$^{252}$Cf, maximum observed flux of fast neutrons in a single
shot can be estimated at the level of up to $1 \times
10^5$~neutrons in $4\pi$~sr. In a shift pulse of neutron radiation
with respect to the x-ray energy neutrons could be estimated under
the assumption about strong correlation x-rays and neutrons. These
estimates give the neutron energy at the level of a few~MeV.

\begin{figure*}
\resizebox{0.90\textwidth}{!}{
\includegraphics{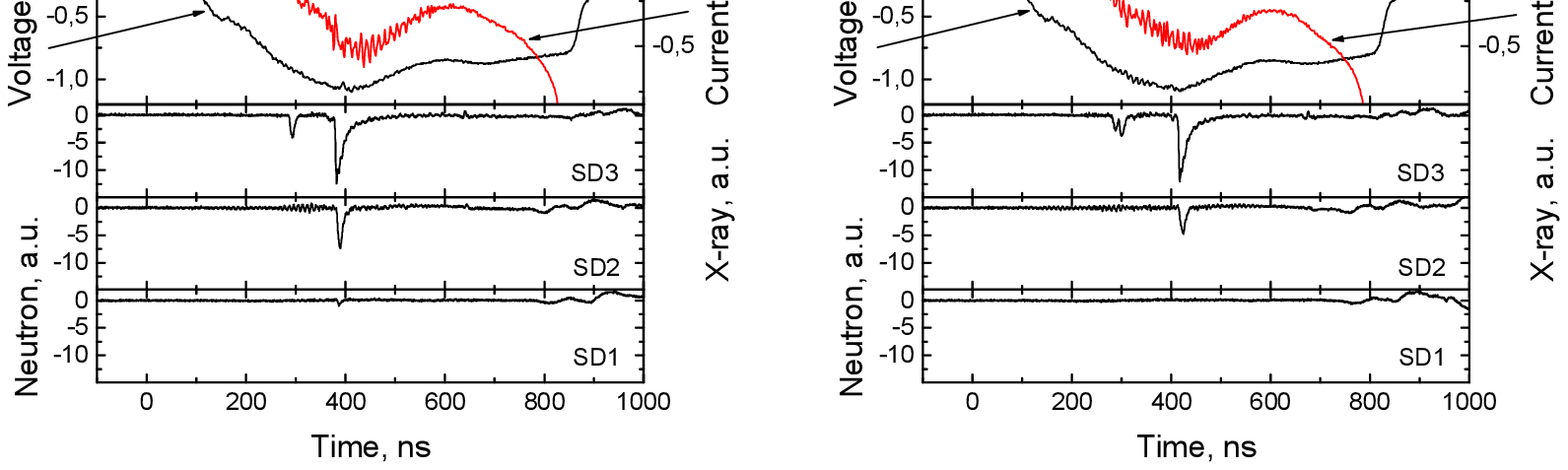}}
\caption{Traces of x-rays of different energy and neutron
radiation relative to voltage and total current traces.}
\label{fig:6}
\end{figure*}

We also used data on multiplicities weakening wide beams of gamma
radiation for 10~cm protection of Pb~\cite{n4}. Fig.~\ref{fig:6} shows the waveform of the pulses detected
by all three scintillation detectors placed in position II (see
Fig.~\ref{fig:1}). In Fig.~\ref{fig:6}a the ratio of the
amplitudes of the signal on the waveform SD3/SD1$\approx 10$,
ratio of the amplitudes SD2/SD1$\approx 6$. With this weakening of
multiplicities correspond to the energy of gamma rays E$_{\gamma}
> 15$~MeV, which is more than an order of magnitude ``applied
voltage'' 1~MeV. Since the emission spectrum is not known, an
estimate gives a lower bound. Of course, such a possibility can
not be rejected completely, however, with high probability we can
assume that this signal (with SD1) is formed by fast neutrons.
Moreover most part of experiments with laboratory discharges give
the energy of gamma rays at level of
150--200~keV~\cite{m1,m2,m3,m4}. At the same time, for another
shot (Fig.~\ref{fig:6}b, which is very similar to
Fig.~\ref{fig:6}a) of this run signal from SD1 is absent, and
ratio of the amplitudes SD3/SD2 $\approx 2.5$. It corresponds to
the energy of gamma rays E$_{\gamma} > 0.5$~MeV in all.

Finally for a detailed study of radiation in the initial
leader-streamer or dark phase of the discharge were carried out a
series of measurements in which especially not a complete
discharge formed. For this purpose, the cathode is installed by
radial rod of adjustable length with a groove which develops on
the outer cylindrical discharge electrode system (reversing
switch). By varying the length of the rod, it is possible to
switch the current from the main (longitudinal) groove on the
discharge gap at a preset time interval, thus interrupting the
discharge development in usual longitudinal direction.

\begin{figure*}
\resizebox{0.90\textwidth}{!}{
\includegraphics{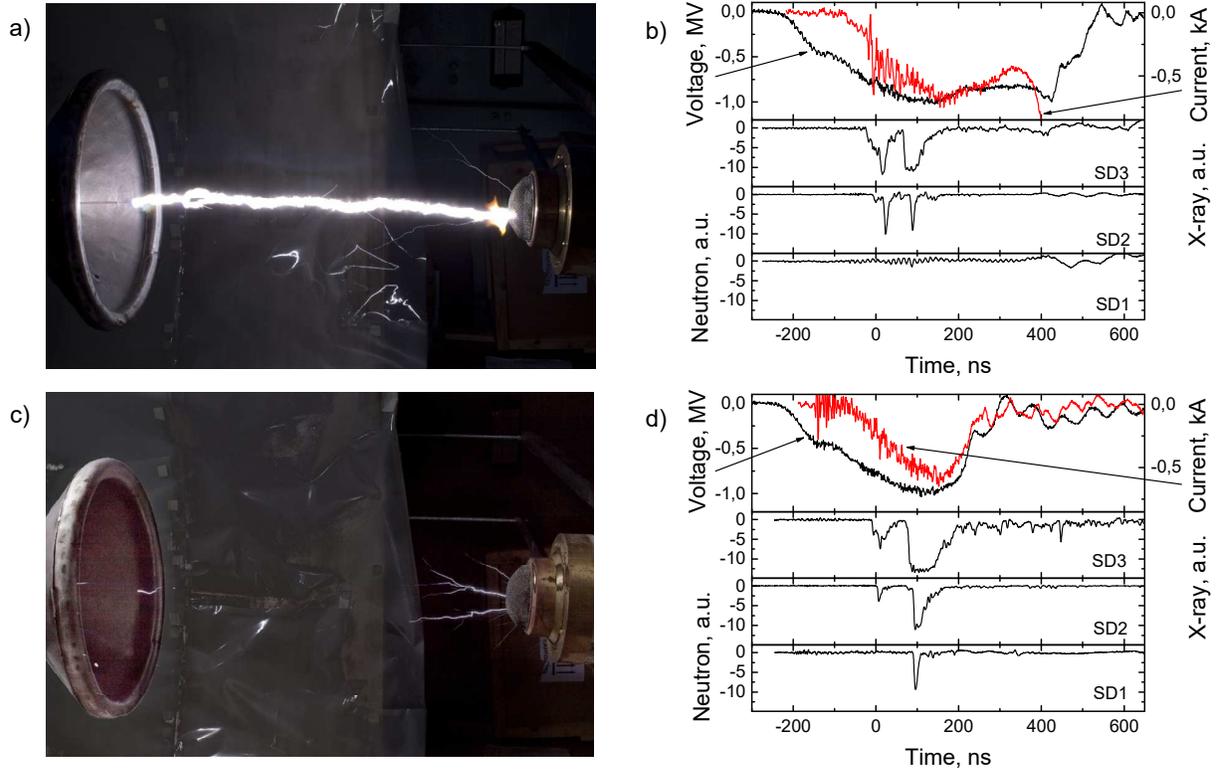}}
\caption{Integral photos and corresponding traces of usual axial
and interrupted discharges.} \label{fig:7}
\end{figure*}

Experiments on the production of integrated image of the initial
phase of atmospheric discharge by interrupting the longitudinal
discharge when current switching to radial discharge were carried
out for two variants of electrode geometry: a cathode and an anode
in the form of a hemisphere with a diameter of 90~mm and a
semispherical anode and cathode needle. In the first case, it was
done 96~shots, of which 34 shots x-rays with photon energies above
10~keV and only 2 pulses of neutrons have been recorded (See
Table~\ref{tab1}). In the second series were only 20~shots. In
each of these shots were recorded x-ray photons with energies
above 100~keV and three shots had pulses of neutrons. Therefore,
we can conclude that the radiation shown in Fig.~\ref{fig:7}d are
formed incomplete discharge in the longitudinal gap configuration
is shown in Fig.~\ref{fig:7}c.

\section{Neutron registration in the atmospheric discharge using $^3$He-detector}

\begin{figure*}
\resizebox{0.90\textwidth}{!}{
\includegraphics{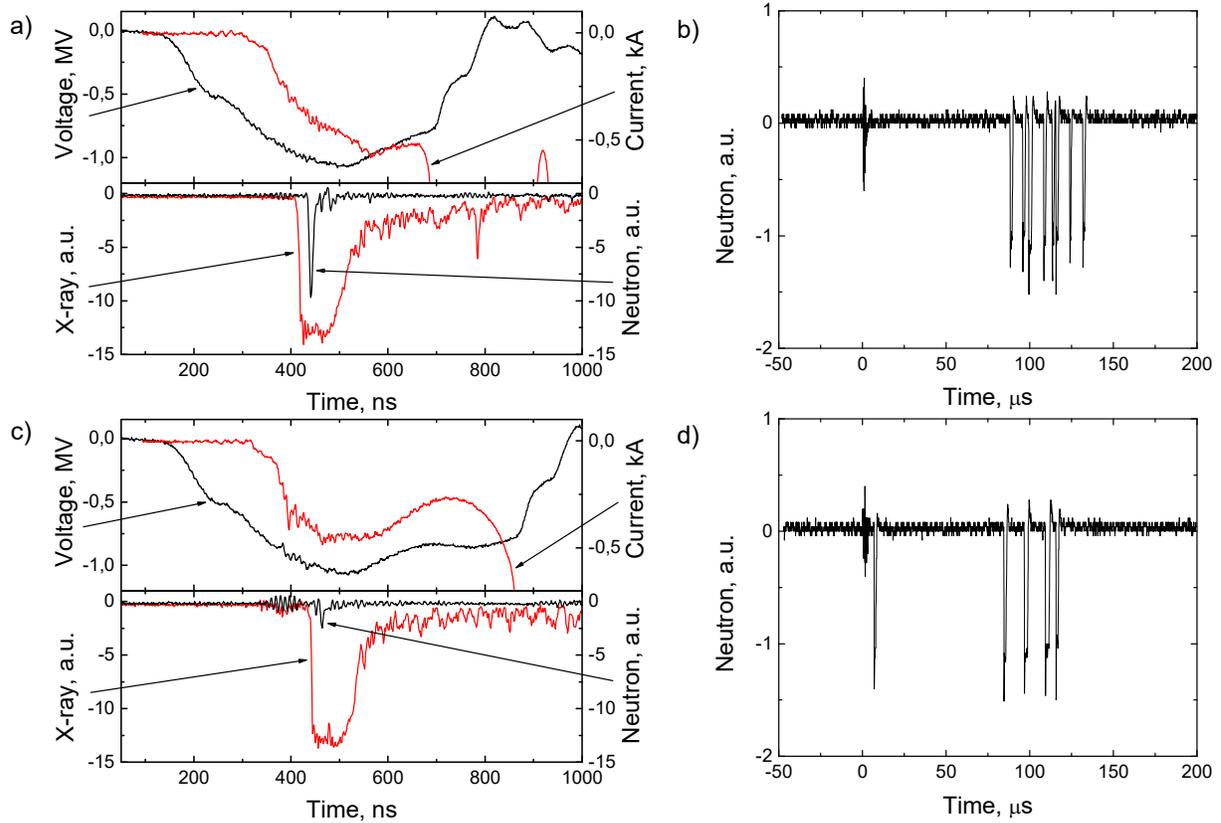}}
\caption{Two illustrations of simultaneous detection of neutrons
with scintillation detectors (left) and $^3$He-detector (right).}
\label{fig:8}
\end{figure*}

To confirm these findings, it was used $^3$He-detector.
Fig.~\ref{fig:8} shows the correlated events, when neutrons
produced per one shot at a time recorded as scintillation
detectors and $^3$He-detector. The left column shows the waveform
from the scintillation detectors, the right column shows the
waveform of the sum signal from the comparator $^3$He-detector
(1~block, 10~tubes).

The first panel shows a case where scintillation detector detects
a neutron signal of large amplitude, which lies inside the x-ray
pulse (left) and a series of neutron pulses recorded $^3$He-
counters (right). The second panel shows a case where there is
neutron signal of small amplitude in the scintillation detector
and the neutrons are also present in the $^3$He-counters.

\begin{figure*}
\resizebox{0.90\textwidth}{!}{
\includegraphics{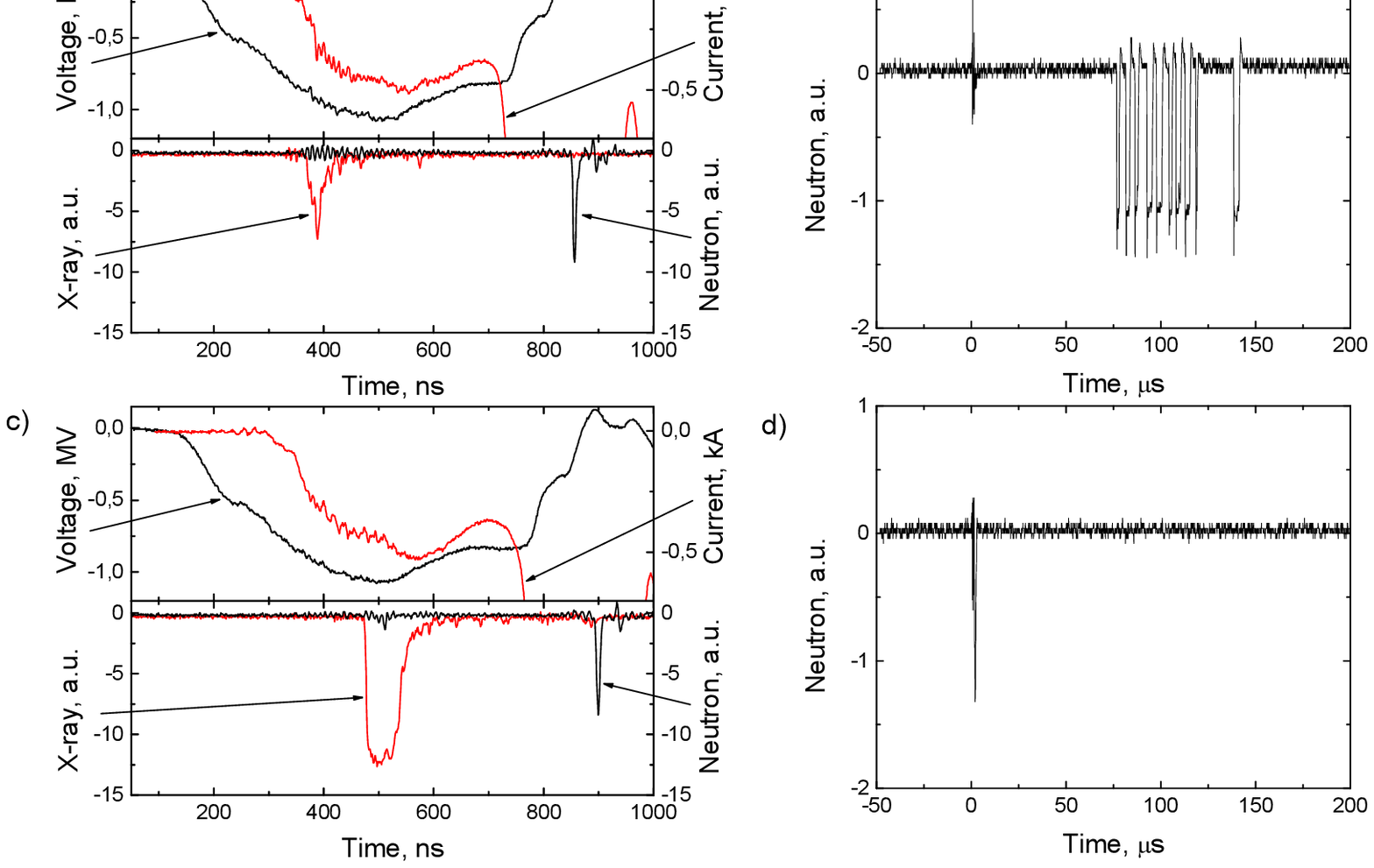}}
\caption{Two illustrations of simultaneous and absent detection of
neutrons with scintillation detectors (left) and $^3$He-detector
(right) for rare time position of neutron pulse.} \label{fig:9}
\end{figure*}

The next two panels are shown in Fig.~\ref{fig:9} for shot cases
in which there is neutron signal of large amplitude in the
scintillation detector at the final stage of the discharge and
recorded neutrons in the $^3$He-counters (first right panel) and
no neutrons case (second right panel).

Table~\ref{tab2} shows data for all the shots in which neutrons
were detected by scintillation detectors and neutron counters.
Experiments have been carried out in configuration semispherical
mesh anode of 90 mm and cathode needle.

\begin{table*}
\caption{\label{tab2}The number of useful signals from the
scintillation detector and $^3$He-counters for the complete series
of shots.}
\begin{ruledtabular}
\begin{tabular}{cccc}
  Total shots&Shots with x-ray ($>10$~keV)&Shots with neutrons on SD (10~cm~Pb)&Shots with neutrons on $^3$He-counters \\ \hline
 340 (100\%)&296 (87\%)&35 (10\%)&44 (13\%) \\
 \\
\end{tabular}
\end{ruledtabular}
\end{table*}

\section{Conclusion}

New data using various combinations of scintillation detectors and
neutron $^3$He counters fully confirmed the neutron radiation in a
laboratory atmospheric discharge.

After analyzing a large discharge statistics, we have established
a strong dependence of the hard x-ray and neutron radiation
appearance on the shape of the electrode. In the configuration of
electrodes in which the anode and cathode is a hemisphere, in 33\%
of the shots have been found hard x-ray radiation and only 1.4\%
of the shots have been registered neutron radiation. In the
configuration of the electrodes in the form of a hemispherical
anode and cathode needle appearance emissions were significantly
more likely to: hard x-ray radiation has appeared in 87\% of the
shots, and neutron emission -- in 15\% of the shots.

In this study, we focused on the temporal structure of neutron
bursts at the time of discharge. Discovered in previous work the
appearance of neutrons in the initial ``dark'' phase of the
discharge was confirmed, however, the temporal structure of
neutron radiation generation was much more diverse. This may
indicate different mechanisms for generating penetrating radiation
during the formation and development of the atmospheric discharge.

Direct evidence for the detection of neutron radiation was
obtained with the use of $^3$He neutron counters and analog
recording of neutron pulses generated at the time of discharge.
The close coincidence of measured and calculated neutron detection
efficiencies of He-counters, allows us to estimate the maximum
value of the flux of neutrons emitted in shots at $1 \times 10^3$
to $5 \times 10^4$ in 4$\pi$~sr, depending on the place of
neutrons generation (near the anode or near the cathode,
respectively). At the same time, scintillation detector placed in
10-cm lead shield does register neutrons, if their energy exceeds
a few hundred~keV. With such small flows is very difficult to talk
about isotropic radiation, so the real overall neutron flux may be
even less. Indirect evidence of this statement can serve the shots
in which the neutrons are recorded only scintillation or $^3$He
detector.


\begin{thebibliography}{00}

\bibitem{dwy} Dwyer J.R., Uman M.A. The physics of lightning. Phys. Rep. V. 534, (2014) 147--241.

\bibitem{sha} G. N. Shah, H. Razdan, G.L. Bhat et al. Nature 313, (1985) 773.
\bibitem{gur} A.V. Gurevich, V. P. Antonova, A. P. Chubenko et al. PRL 108 (2012) 125001.

\bibitem{gur2} A.V. Gurevich, V. P. Antonova, A. P. Chubenko et al. Atmospheric Research 164 (2015) 339.

\bibitem{chi} A. Chilingarian, A. Daryan, K. Arakelyan et al., Phys. Rev. D 82, (2010) 043009.

\bibitem{chi2} A. Chilingarian, N. Bostanjyan, and L. Vanyan, Phys. Rev. D 85, (2012) 085017.

\bibitem{tsu} H. Tsuchiya, K. Hibino, K. Kawata et al., Phys. Rev. D85, (2012) 092006.

\bibitem{kuzh} V. M. Kuzhevsky, Bulletin of Moscow University: Physics, Astronomy 5, (2004) 14.

\bibitem{koz} V. I. Kozlov, V. A. Mullayarov, S. A. Starodubtsev, and A. A.
Toropov, J. Phys. Conf. Ser. 409, (2013) 012210.

\bibitem{aga} A. V. Agafonov, A. V. Bagulya, O. D. Dalkarov et al., PRL 111, (2013) 115003.

\bibitem{bab} L. P. Babich, Phys.Rev. C, 92, (2015) 044602.

\bibitem{n1} A.V. Agafonov, A.V. Oginov and K.V. Shpakov K.V. Physics of Particles and Nuclei Letters. V. 9, (2012), 380--383.

\bibitem{n2} G.A. Mesyats, High current pulsed electron beam technology, Nauka, Novosibirsk (1983) 130 (in Russian).

\bibitem{geant4} Geant4 Collaboration, NIM A, 506, (2003) 250.

\bibitem{n4} I. K. Kikoin, Tables of Physical Values, Atomizdat, Moscow (1976) (in Russian).

\bibitem{m1} J. R. Dwyer, H. K. Rassoul, Z. Saleh et al., Geophys. Res. Lett. 32, (2005) L20809.

\bibitem{m2} J. R. Dwyer , Z. Saleh, H. K. Rassoul et al., J. Geophys. Res., Vol. 113, (2008) D23207.

\bibitem{m3} V. March, J. Montany$\grave{\mbox{a}}$, Geophys. Res. Lett., 37, (2010) L19801.

\bibitem{m4} P. O. Kochkin, C. V. Nguyen, A. P. J. van Deursen et al., J. Phys. D: Appl. Phys. 48, (2015) 025205.

\end{thebibliography}
\end{document}